\documentclass[conference]{IEEEtran}
\usepackage{aas_macros}
\usepackage{cite}
\usepackage{amsmath,amssymb,amsfonts}
\usepackage{algorithmic}
\usepackage{graphicx}
\usepackage{textcomp}
\usepackage{xcolor}
\usepackage{dsfont}
\usepackage{url}
\usepackage[caption=false]{subfig}
\begin{document}

\title{Morphological Classification of Radio Galaxies using Semi-Supervised Group Equivariant CNNs}
\author{\IEEEauthorblockN{Mir Sazzat Hossain\IEEEauthorrefmark{1},
Sugandha Roy\IEEEauthorrefmark{1},
K. M. B. Asad\IEEEauthorrefmark{1}\IEEEauthorrefmark{2}\IEEEauthorrefmark{3}, 
Arshad Momen\IEEEauthorrefmark{1}\IEEEauthorrefmark{2}, 
Amin Ahsan Ali\IEEEauthorrefmark{1},\\
M Ashraful Amin\IEEEauthorrefmark{1} and
A. K. M. Mahbubur Rahman\IEEEauthorrefmark{1}}
\IEEEauthorblockA{\IEEEauthorrefmark{1}Center for Computational \& Data Sciences, Independent University, Bangladesh}
\IEEEauthorblockA{\IEEEauthorrefmark{2}Department of Physical Sciences, Independent University, Bangladesh}
\IEEEauthorblockA{\IEEEauthorrefmark{3}Astronomy and Radio Research Group, SETS, Independent University, Bangladesh}
\IEEEauthorblockA{{\tt\small \{mirsazzathossain, sugandha.roy08\}@gmail.com, \{kasad, arshad, aminali, aminmdashraful,}\\
    {\tt\small akmmrahman\}@iub.edu.bd}}
}

\maketitle

\begin{abstract}
Out of the estimated few trillion galaxies, only around a million have been detected through radio frequencies, and only a tiny fraction, approximately a thousand, have been manually classified.
We have addressed this disparity between labeled and unlabeled images of radio galaxies by employing a semi-supervised learning approach to classify them into the known Fanaroff-Riley Type I (FRI) and Type II (FRII) categories.
A Group Equivariant Convolutional Neural Network (G-CNN) was used as an encoder of the state-of-the-art self-supervised methods SimCLR (A Simple Framework for Contrastive Learning of Visual Representations) and BYOL (Bootstrap Your Own Latent).
The G-CNN preserves the equivariance for the Euclidean Group E(2), enabling it to effectively learn the representation of globally oriented feature maps.
After representation learning, we trained a fully-connected classifier and fine-tuned the trained encoder with labeled data.
Our findings demonstrate that our semi-supervised approach outperforms existing state-of-the-art methods across several metrics, including cluster quality, convergence rate, accuracy, precision, recall, and the F1-score. 
Moreover, statistical significance testing via a t-test revealed that our method surpasses the performance of a fully supervised G-CNN. 
This study emphasizes the importance of semi-supervised learning in radio galaxy classification, where labeled data are still scarce, but the prospects for discovery are immense.
\end{abstract}


\section{Introduction} \label{introduction}
Radio astronomy has given us unprecedented access to explore galaxies far beyond what is visible to the naked eye.
The Karl G. Jansky Very Large Array (VLA), a U.S.-based radio telescope array, has imaged a significant number of radio galaxies that are exceptionally bright at radio frequencies.
These galaxies display a variety of morphologies, such as Fanaroff-Riley (FR) Types I and II \cite{fanaroff74}, head-tailed  \cite{sasmal2022}, ringlike  \cite{proctor11} and x-shaped  \cite{leahy92}, among others.
Classifying these morphologies is crucial for our understanding of the universe. However, manual classification has become nearly impossible because the number of detected galaxies is exploding with the improvement of telescope technologies. As such, the scientific community is exploring artificial intelligence, particularly machine learning techniques, to address this challenge.

In this work, we present a semi-supervised approach for classifying radio galaxies. Our proposed method has two steps: task-agnostic self-supervised learning and task-specific fine-tuning.
Initially, we used the self-supervised techniques SimCLR (A Simple Framework for Contrastive Learning of Visual Representations) \cite{chen20} and BYOL (Bootstrap Your Own Latent) \cite{grill20} to learn representations from a large unlabeled dataset.
We then fine-tuned these representations using a smaller labeled dataset, specifically for classifying the galaxies into either FRI or FRII categories.
This semi-supervised approach uses losses that encourage the encoder to extract robust representations from large amounts of unlabeled data.

Galaxy images can have random orientations. In order to address this challenge, we have modified the encoders of the self-supervised models so that the network is equivariant to different isometries, such as translation, rotation, and mirror reflection.
Specifically, we have used a D16 (Dihedral group with 16 rotations) equivariant CNN proposed by Scaife and Porter  \cite{scaife21} as a feature extractor in SimCLR and BYOL, which ensures that the extracted features are equivariant to the isometries.
After representation learning, we fine-tuned the D16 equivalent CNN with labeled data and trained fully connected layers to classify the FR-type radio galaxies.

Using a semi-supervised approach involving self-supervised learning followed by supervised fine-tuning, we devised a strategy that effectively utilizes a large number of unlabeled radio galaxy images and successfully tackles the random orientations of galaxies.

In summary, our paper presents the following novel contributions to the classification of radio galaxies:

\begin{itemize}
    \item We have used Group Equivariant Convolutional Neural Network (G-CNN) as an encoder of SimCLR and BYOL for the first time.
    \item Our semi-supervised approach outperforms the state-of-the-art methods in FR classification.
    \item We have found evidence that representation learning from a large dataset of unlabeled data is beneficial.
    \item Our method can be successful in tasks where the labelled dataset is very small compared to the available unlabelled data.
\end{itemize}

This paper is structured to provide a clear and organized description of our work on radio galaxy classification.
Section \ref{sci} gives a brief overview of radio galaxies and the need for using machine learning in classifying them.
Section \ref{lit} describes the relevant previous works of radio galaxy classification using machine learning.
In Section \ref{background},  we discuss the relevant parts of the specific models used in our work.
Section \ref{method} presents our proposed model and its unique training strategy.
Section \ref{data} covers the datasets used in our study.
Our experimental setup and results are presented in Sections \ref{setup} and \ref{results}, respectively.
Finally, in Section \ref{conclusion}, we summarize our key findings and conclusions.
We emphasize the superior performance of our radio galaxy classification strategy compared to other existing models.

\section{Scientific background} \label{sci}

The cosmos teems with a diverse array of galaxies, but due to technological limitations, astronomers have only been able to observe a fraction of them. 
These observations are made at various frequencies such as optical (around 400--800 THz), infrared (around 1--30 THz) and radio (few MHz -- few GHz). 
Galaxies appear to be different at different frequencies. 
Radio galaxies, which emit a significant amount of radiation at radio wavelengths, have been categorized into various morphologies.

\begin{figure}
    \centering
    \subfloat[FRI\label{fig:fri}]{\includegraphics[width=0.48\linewidth]{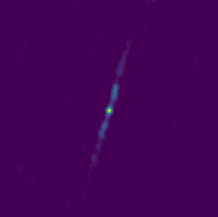}}
    \hfill
    \subfloat[FRII\label{fig:frii}]{\includegraphics[width=0.48\linewidth]{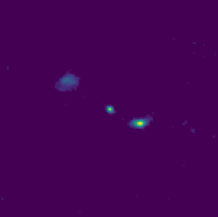}}
    \caption{Examples of Fanaroff-Riley galaxies. (a) FRI galaxy, characterized by peak radio emission near the core and darker edges of the lobes. (b) FRII galaxy, characterized by peak radio emission at the edge of the lobes far from the core.}
  \label{fig1} 
\end{figure}

The Fanaroff-Riley (FR) categorization, based on the physical characteristics of a galaxy, is crucial for understanding the underlying science behind the formation and development of galaxies.
Typical radio galaxies have a supermassive black hole in their core that actively accretes gas and stars along its equator and, as a result, ejects multiple jets through its poles.	
Consequently, radio images of a galaxy usually consist of a core and multiple lobes. FRI galaxies, as depicted in \figurename{\ref{fig:fri}}, have the peak of their radio emission near the core, and the lobes have a darker edge than their core.
On the other hand, FRII galaxies, as shown in \figurename{\ref{fig:frii}}, have peak emissions at the edge of the lobes far from the core.

Despite many detected radio galaxies, a mere fraction, approximately one thousand, have been manually classified as either FRI or FRII  \cite{miraghaei17}.
However, with the imminent operation of the Square Kilometre Array (SKA), the most significant scientific facility ever constructed, and its pathfinder telescopes, sensitivity and survey speed will increase significantly.
This will enable the imaging of tens of thousands of radio sources in a matter of hours, a task that would have taken months to accomplish only a few years ago  \cite{galvin20}.
The sheer volume of data generated by the SKA requires using artificial intelligence to supplement human analysis.
Hence, the radio astronomy and computing communities are actively engaged in the application of machine-learning techniques to classify radio galaxies.

\section{Related previous works} \label{lit}

Images provided by the Faint Images of the Radio Sky at Twenty-Centimeters (FIRST) survey have been crucial for researchers in radio galaxy classification. 
This survey, conducted using the VLA radio telescope, offers some of the most detailed observations of the radio sky, predominantly at a frequency of 1.4 GHz. 
In the quest to classify these complex radio galaxy images, various machine-learning techniques have been adopted. 
Such as using a Convolutional Neural Network (CNN) in the form of a slightly modified AlexNet as done by Aniyan and Thorat \cite{aniyan17}, or using a CNN with three convolution layers as demonstrated by Alhassan et al. \cite{alhassan18}. 
Wu et al. \cite{wu19} employed Faster Region-based Convolutional Neural Networks (Faster R-CNN) to classify galaxies into six categories based on the number of components and peaks within a source. 
Tang et al. \cite{tang19} adopted a 13-layer CNN and investigated the identification ability of classification networks on cross-survey data using Transfer Learning. Since deep networks provide better performance in classification, researchers are increasingly utilizing such networks to classify radio galaxies.

One of the main challenges in classifying radio galaxies using CNNs is the need for the network to be equivariant to different isometries such as translation, rotation, and mirror reflection. 
Since radio galaxies can have different orientations, CNN needs to be able to classify them regardless of their position or symmetry. 
While CNNs are typically translation-invariant, they must also be able to handle other types of isometries.
Previous methods have attempted to address this issue by augmenting the data with rotated images. 
However, this approach has proven ineffective as the CNN may learn multiple copies of the same kernel differently. Scaife and Porter \cite{scaife21} addressed this issue by using G-CNNs \cite{cohen16}, which can preserve equivariance for all isometries, allowing for a more accurate classification of radio galaxies.

Another key challenge in classifying radio galaxies is leveraging the vast amount of unlabeled data. 
A significant portion of available radio galaxy data needs to be labeled, making it difficult for supervised methods to utilize this potential wealth of information fully. 
Using this unlabeled data could increase the model's generalization ability, but the absence of labels makes it challenging to derive meaningful insights directly. 
Self-supervised learning techniques can, however, be used to extract useful information from these unlabeled data and apply it in a task-specific manner to improve the performance of our models. 
This approach effectively addresses the scarcity of labelled data in radio galaxy classification, as researchers have already begun to explore this method.
For example, Ma et al. \cite{ma19} developed a CNN-based autoencoder named Morphological Classification of Radio Galaxy Network (MCRGNet), consisting of an encoder and a decoder block. 
First, they trained the MCRGNet (Morphological Classification of Radio Galaxy Network) autoencoder with unlabeled data and then fine-tuned the pre-trained encoder with labelled data to classify radio morphologies. 
This method performed significantly better than traditional supervised classifications \cite{aniyan17,alhassan18}.

A later work by Ma et al. \cite{ma19c} used a VGG16 network as an autoencoder and achieved improved performance. 
However, more than a simple reconstruction error is needed to learn invariant representations under various noises and transformations. 
Slijepcevic et al. \cite{slijepcevic22} used the semi-supervised learning technique FixMatch, which combines labelled and unlabeled data to improve model performance.
This method uses a self-supervised approach to generate pseudo-labels for unlabeled data, which are then used to fine-tune the model with labelled data. This helps to extract useful information from large amounts of unlabeled data, resulting in improved model performance.
However, the improvement is still relatively small compared to the baseline \cite{scaife21} with fewer labels \cite{slijepcevic22}.


\section{Related deep-learning architectures} \label{background}

In the previous section, we discussed various techniques used for radio galaxy classification, highlighting the challenges and limitations in the field.
Motivated by these, our approach utilizes the G-CNN, and self-supervised learning techniques such as SimCLR \cite{chen20} and BYOL \cite{grill20}.
These methods form the cornerstone of our proposed technique, which adopts the `unsupervised pretrain, supervised fine-tune' paradigm.
This section outlines the pertinent features of these deep learning architectures, setting the groundwork for the detailed discussion of our proposed method in Section \ref{method}.

\subsection{Group Equivariant CNN (G-CNN)}

G-CNN \cite{cohen16} is a specialized neural network architecture that possesses the capability of being equivariant to symmetries present in the input data, such as translations, rotations, and reflections. 
This feature ensures that the network produces the same output when the input data is transformed in a particular way in accordance with the symmetries of a group.

G-CNN employs the theory of group representations to establish the convolution operation.
This is achieved by substituting the traditional convolution operation with a group convolution operation, mathematically represented as

\begin{equation}
(f*\phi)(g) = \sum_{h \in G} f(h)\phi(h^{-1}g),
\label{eq:gcnn}
\end{equation}

\noindent
where $f$ and $\phi$ are the input and kernel functions, respectively, $G$ is the group of symmetries while $g$ and $h$ are its elements.

\subsection{A Simple Framework for Contrastive Learning of Visual Representation (SimCLR)}

SimCLR  \cite{chen20} is a self-supervised learning method that aims to learn representations of the input data by maximizing the agreement between two different views of the same data.
The method utilizes the concept of contrastive learning, which is based on the idea of comparing the similarity between two different data samples.

SimCLR uses two neural networks, an encoder and a projection head.
The encoder is trained to map the input data to a feature space, while the projection head is trained to map the features back to the input space.
The two networks are trained together using a contrastive loss function, defined as the negative log-likelihood of the correct pair of views, mathematically

\begin{equation}
\mathcal{L}_{\text{SimCLR}} = -\log \frac{\exp(\text{sim}(z_i,z_j)/\tau)}{\sum_{k=1}^{2n} \mathds{1}_{[ k\ne i] } \exp(\text{sim}(z_i,z_k)/\tau)},
\label{eq:simclr}
\end{equation}

\noindent
where $z_i$ and $z_j$ are the features of the two views of the same data sample, $z_k$ are the features of a different sample; the dataset has $n$ number of samples in total.
The indicator function $\mathds{1}_{[ k\ne i] }$ equals 1 when $k \neq i$, ensuring the sum considers only negative pairs (different sample) and excludes the current positive pair (same data sample).
$\text{sim}(\cdot)$ is a similarity function, such as cosine similarity, and $\tau$ is a temperature parameter that controls the concentration of the distribution.
The objective of the loss function is to maximize the agreement between the two views of the same sample and minimize the agreement between different samples.

\subsection{Bootstrap Your Own Latent (BYOL)}

BYOL \cite{grill20} is a self-supervised learning method that utilizes the idea of ``global contrastive learning'' to learn representations of input data.
The method simultaneously trains two neural networks, an online network and a target network.
The online network, parametrized by $\theta$, is used to generate predictions of the input data.
The target network, parametrized by $\xi$, is used to generate predictions of the predictions of the online network.
The goal of BYOL is to learn similar representations between the two networks, which are, in turn, learned from the input data.

The training process of BYOL is guided by a contrastive loss function, defined as the mean squared error (MSE) between the feature representations of the online network and the target network.
Mathematically, this loss function is represented as

\begin{equation}
\mathcal{L}_{\theta, \xi} \triangleq \left\lVert \overline{q}_{\theta}(z_{\theta}) - \overline{z}'_{\xi}\right\rVert^{2}_{2} = 2 - 2 \cdot \frac{\langle q_{\theta}(z_{\theta}), z'_{\xi}\rangle}{\left\lVert q_{\theta}(z_{\theta})\right\rVert_{2} \cdot \left\lVert z_{\xi}' \right\rVert_{2}},
\label{eq:byol}
\end{equation}

\noindent
where $q_{\theta}(z_{\theta})$ is the feature representation of the input data generated by the online network, and $z'_{\xi}$ is the feature representation of the input data generated by the target network.
The goal of the loss function is to encourage the online network to learn representations similar to the representations of the target network.

The main advantage of BYOL is that it does not require explicit data augmentation or negative samples during training; it only requires the data to learn the representations.

\section{Our proposed method}\label{method}

\begin{figure*}
    \centering
    \subfloat[SimCLR\label{fig:simclr}]{\includegraphics[width=0.48\linewidth]{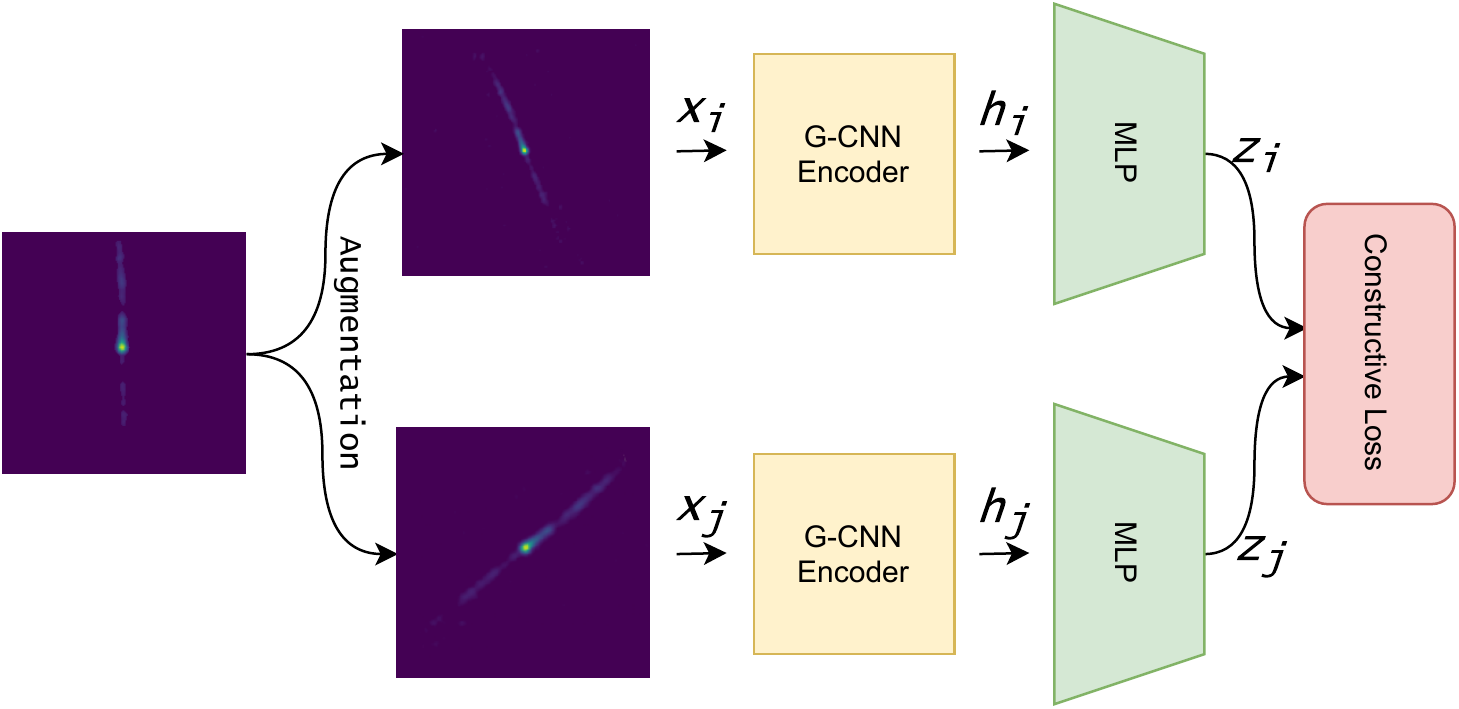}}
    \hfill
    \subfloat[BYOL\label{fig:byol}]{\includegraphics[width=0.48\linewidth]{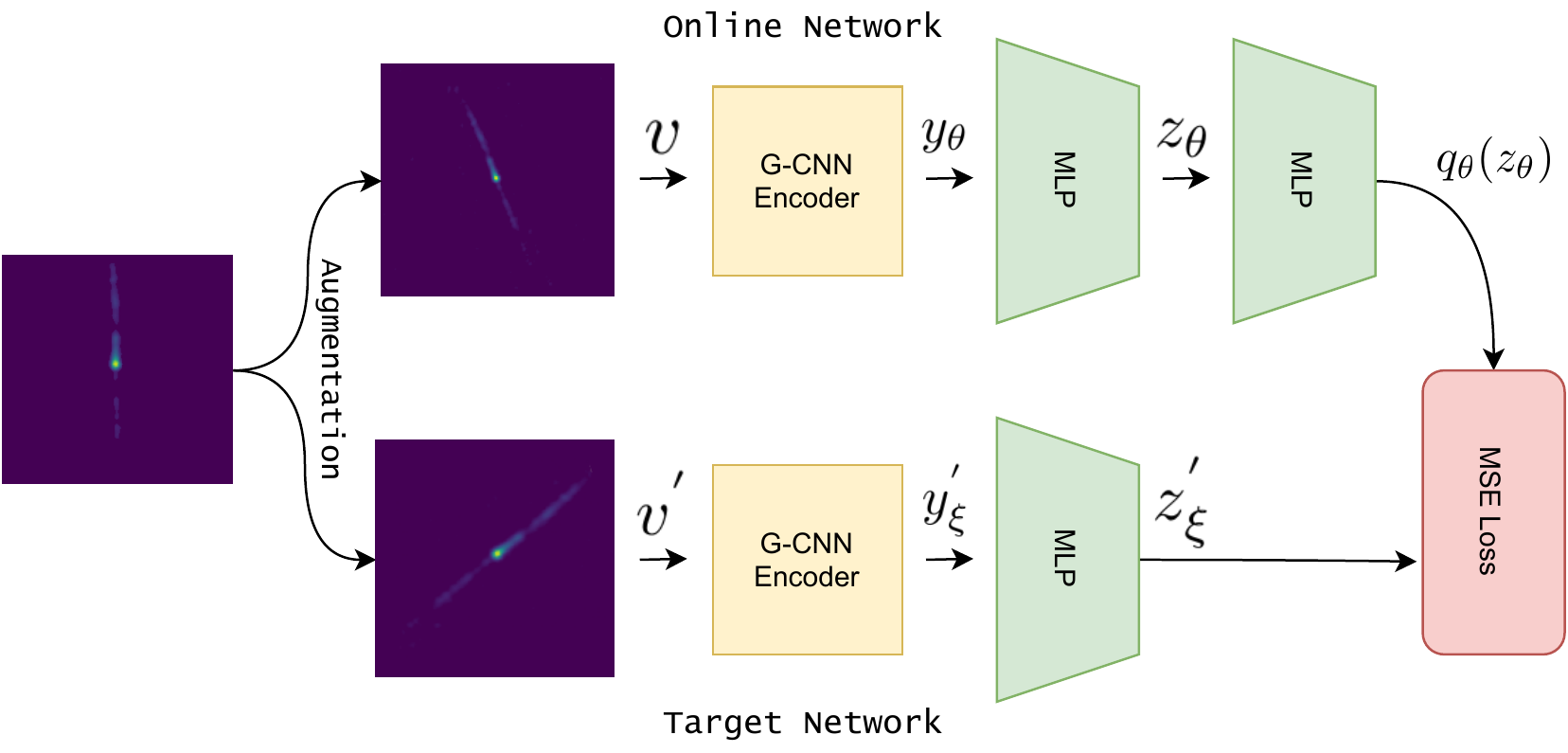}}
    \caption{Illustration of (a) SimCLR and (b) BYOL architectures used for self-supervised representation learning from an unlabeled dataset. The data augmentation strategy for radio galaxy images has been modified by setting bounds for aspect ratio, adding an extra random rotation, and replacing the ResNet-50 feature extractor with an E(2)-Equivariant Steerable G-CNN for improved performance on the downstream radio galaxy classification task.}
    \label{fig:ssl}
\end{figure*}

In this paper, we propose a semi-supervised approach for radio galaxy classification.
Our method consists of two main steps: task-agnostic self-supervised learning and task-specific fine-tuning.
In the first step, we use the state-of-the-art self-supervised representation learning techniques SimCLR and BYOL to learn robust representations from a large amount of unlabeled data. 
In the second step, we fine-tune these representations using a small amount of labelled data for the specific task of radio galaxy classification.
The use of a semi-supervised approach allows us to leverage a large amount of unlabeled data while still achieving high performance in task-specific classification.
We have described these two steps in more detail and provided an in-depth analysis of our proposed method below.

\subsection{Task-agnostic self-supervised learning}
In this step, we employ the BYOL and SimCLR methods to learn representations of the input data using an unlabelled dataset.
Fig. \ref{fig:ssl} illustrates the SimCLR and BYOL architectures adopted for self-supervised representation learning from the unlabeled dataset.

SimCLR, depicted in Fig. \ref{fig:simclr}, utilizes an encoder $f(\cdot)$ that extracts features from different augmented views $x_i$ and $x_j$ of the same image $x$.
These augmented views generated from the same image are considered positive pairs of samples.
Assuming a batch size of $N$, the other $2(N - 1)$ augmented examples are treated as negative pairs of samples.
The representations, $h_i = f(x_i)$ and $h_j = f(x_j)$, are then mapped to the latent space using a projection head $g(\cdot)$, which is a Multi-Layer Perceptron (MLP) with one hidden layer.
The contrastive loss (Eq. \ref{eq:simclr}) is applied to the mapped representations $z_i = g(h_i)$ and $z_j = g(h_j )$.

BYOL, illustrated in Fig. \ref{fig:byol}, creates two augmented views, $v$ and $v'$, of an image.
The feature extractor of the online network outputs a representation $y_{\theta} \triangleq f_{\theta}(v)$, and the projection head provides a projection $z_\theta \triangleq g_\theta(y_\theta)$ from the first augmented view $v$.
From the second augmented view $v'$, the target network produces $y'_{\xi} \triangleq f_{\xi}(v')$ and $z'_{\xi} \triangleq g_{\xi}(y'_{\xi})$.
An additional MLP, serving as a predictor, is used in the online network. 
This MLP maps the online network's projection $z_\theta$ to a space corresponding to the target network's output. Here, $q_{\theta}(z_{\theta})$ seeks to approximate the output of the target network's projection, $z'_{\xi}$. 
Minimizing the MSE loss (Eq. \ref{eq:byol}) between the $l_2$-normalized $q_{\theta}(z_{\theta})$ and $z'_\xi$ serves to enhance the alignment between $q_{\theta}(z_{\theta})$ and $z'_\xi$.

In the case of SimCLR and BYOL, the authors used a combination of cropping, colour distortion, and Gaussian blur as data augmentation techniques.
They found that this combination of augmentations improved the performance of these self-supervised models.
However, the default augmentations used in SimCLR and BYOL were unsuitable for radio galaxy images because the galaxies occupy a very small portion of the images.
Random cropping and resizing could result in many empty patches, negatively impacting the model's training.

Our contribution to this work is redesigning the data augmentation policy for radio galaxy images.
We retained the colour distortion and Gaussian blur augmentations, set lower and upper bounds for the random aspect ratio of the crop to 0.8 and 1, respectively, and added an extra $360^{\circ}$ random rotation to the augmentations.
This helped to ensure that most of the cropped patches contained the galaxy while increasing the diversity of the training data.
By redesigning the data augmentation strategy, we improved the quality of the learned representations and achieved better results.

In addition to redesigning the data augmentation strategy, we have also made another significant contribution to this work by replacing the ResNet-50 feature extractor of both BYOL and SimCLR with the E(2)-Equivariant Steerable G-CNN, specifically the D16 Steerable G-CNN.
This network contains 2 group convolution layers with kernel size $5\times 5$, each followed by a ReLU activation and max pooling.
The output is then passed through a linear layer that maps the 20736-dimensional feature maps to a feature map of dimension 2048.
This feature map is then used as input for the MLP projection head.
This change allows the network to be equivariant to different isometries, enabling it to classify radio galaxies with different orientations, regardless of their position or symmetry.

\subsection{Task-specific fine-tuning}

In order to optimize the model for the task of FR classification, we perform task-specific fine-tuning on our labelled dataset next. This step involves modifying the pre-trained encoder obtained through task-agnostic self-supervised learning to better suit our labelled data's characteristics.

 \begin{figure}[!ht]
    \centering
    \includegraphics[width=0.8\linewidth]{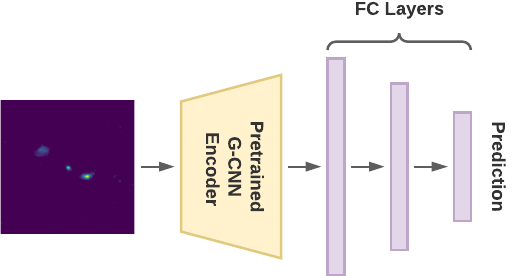}
    \caption{Illustration of the fine-tuning step for FR classification. The pre-trained encoder from task-agnostic self-supervised learning is modified by replacing the last fully connected layer with three new fully connected layers, allowing the model to adapt to the specific task of FR classification.}
    \label{fig:downstream}
\end{figure}

As shown in Fig. \ref{fig:downstream}, the final fully connected layer of the encoder is replaced with a new architecture that consists of three fully connected linear layers. 
The first linear layer transforms the 20736-dimensional feature maps into a 120-dimensional feature map, the second linear layer maps the 120-dimensional feature maps to an 84-dimensional feature map, and finally, the third linear layer reduces it to a 2-dimensional feature map. 
We applied ReLU activation functions after the first and second linear layers and a dropout layer after the second linear layer.

This modification allows the network to adapt to the specific task of FR classification and enhances its ability to recognize the distinctive characteristics of the labelled data. Fine-tuning uses a supervised learning approach and cross-entropy loss function on the labelled data. 
We aim to adapt the model better to recognize the unique characteristics of our labeled data.


\section{Datasets}\label{data}
Throughout this work, we utilized radio galaxy images from the Karl G. Jansky VLA as part of the FIRST survey.
The survey covered a vast area of 10,000 square degrees of the sky, producing images with a resolution of 5 arcsec and a sensitivity of $0.15$ mJy (milli-jansky; $1$ Jy = $10^{-26}$ W m$^{-2}$ Hz$^{-1}$).
Each pixel in the final images corresponds to an angular size of 1.8 arcsec.
We used pre-processed FIRST survey images to perform representation learning and radio galaxy classification.

For representation learning, we used the Radio Galaxy Zoo (RGZ) dataset  \cite{banfield15}, pre-processed by Wu et al.  \cite{wu19}\footnote{\url{https://github.com/chenwuperth/rgz_rcnn}}. 
The original dataset consisted of 11,678 images\footnote{\url{https://cloudstor.aarnet.edu.au/plus/s/agKNekOJK87hOh0}}, of which we selected approximately 9,700 images where only one galaxy was visible.
Each image originally had dimensions of $132\times 132$ pixels, subsequently padded with zeros to $150\times 150$ pixels. 
This unlabeled dataset will be referred to as Dataset-U in the following sections.

For the fine-tuning step with a labeled dataset, we have used FIRST images labeled by Miraghaei and Best  \cite{miraghaei17} and pre-processed for FR classification by Porter  \cite{porter20}. The dataset, referred to as the MiraBest Batched Dataset, contains 1256 images, each of which is assigned a three-digit numerical identifier.
The first digit of this identifier denotes the source class: 1 for FRI, 2 for FRII, and 3 for the hybrid sources.
The second digit indicates the level of confidence in the classification: 0 for confidently classified and 1 for uncertainly classified.
The third digit denotes the morphology of the source: 0 for standard morphology, 1 for double-double, 2 for wide-angled tail, 3 for diffuse source, and 4 for head-tail.
The meaning of each digit is summarized in Table \ref{tab:mirabest-num}.

\begin{table}
    \renewcommand{\arraystretch}{1.3}
    \caption{Numerical Identifiers of MiraBest Batched Dataset}
    \label{tab:mirabest-num}
    \centering
    \begin{tabular}{lll}
         \hline
         Digit 1 & Digit 2 & Digit 3 \\
         \hline
         1 - FRI &  0 - Confident & 0 - Standard\\
         2 - FRII & 1 - Uncertain & 1 - Double-double\\
         3 - Hybrid & & 2 - Wide-angle Tail \\
           & & 3 - Diffuse \\
         & & 4 - Head-tail \\
         \hline
    \end{tabular}
\end{table}

In our work, we have only utilized confident sources from the MiraBest Batched Dataset, specifically those with numerical identifications 100, 102, 104, 200, or 201.
We named this labeled data Dataset-F which was utilized for training and testing purposes.
Like Dataset-U, each image in Dataset-F has a $150\times 150$ pixels resolution. 
The breakdown of source counts for FRI and FRII sources in the training and testing sets of Dataset-F can be found in Table \ref{tab:dataset-f}.

\begin{table}
    \renewcommand{\arraystretch}{1.3}
    \caption{Source counts for FRI and FRII sources in the labelled dataset (Dataset-F)}
    \label{tab:dataset-f}
    \centering
    \begin{tabular}{l|ll|l}
        \hline
         & \multicolumn{2}{ c| }{MiraBest} \\
         \cline{2-3}
        Morphology  & Train & Test & Total \\
        \hline
        FRI       & 348 & 49 & 397 \\
        FRII      & 381 & 55 & 436 \\
        \hline
        Total     & 729 & 104 & 833 \\
        \hline
    \end{tabular}
\end{table}


\section{Experimental analysis}\label{experiment}
This section presents our experimental setup and provides a detailed analysis of the results obtained through our semi-supervised approach for radio galaxy classification.

\begin{figure*}
    \centering
    \subfloat[BYOL\label{fig:tsne_finetuned_byol}]{\includegraphics[width=0.32\linewidth]{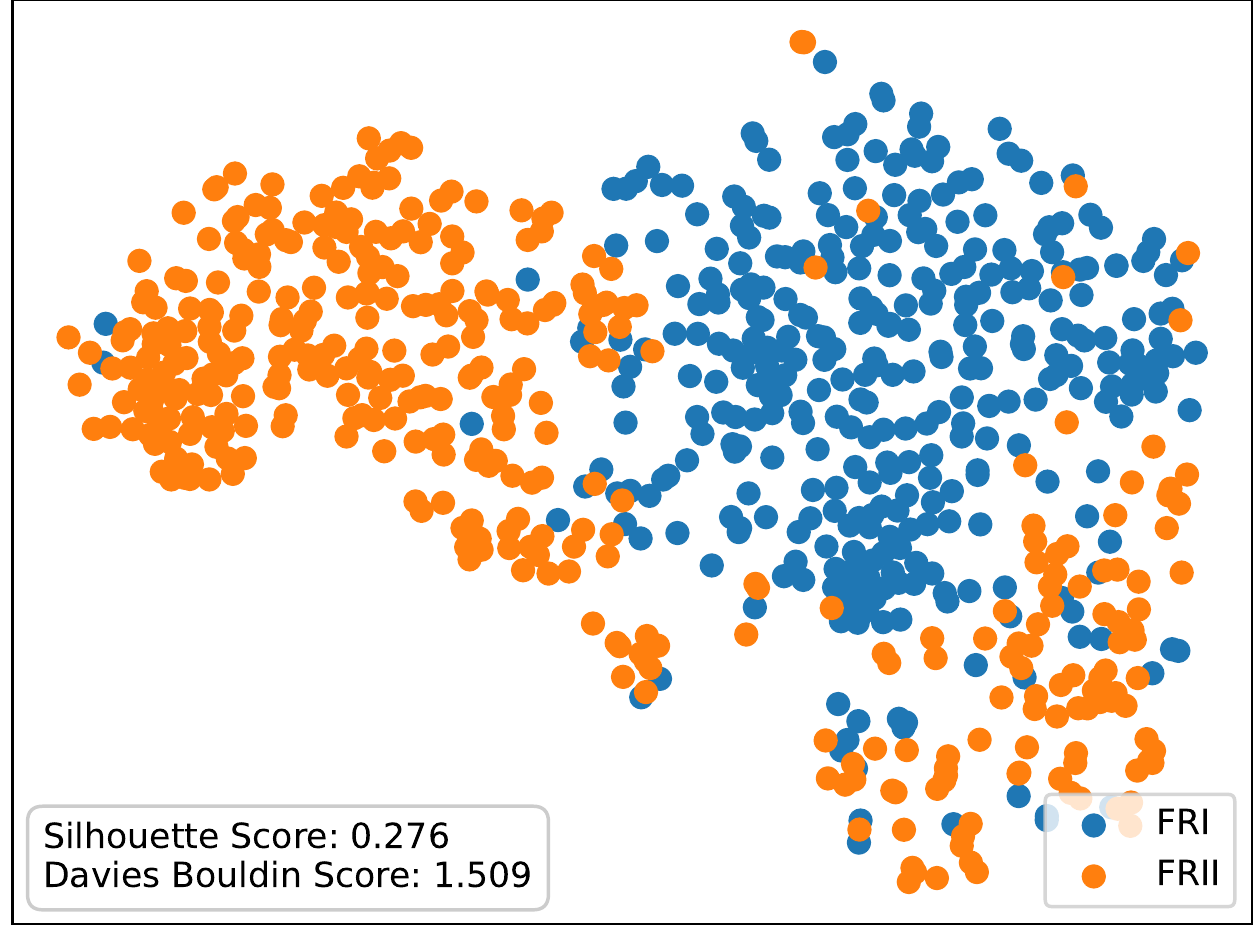}}
    \hfill
    \subfloat[SimCLR\label{fig:tsne_finetuned_simclr}]{\includegraphics[width=0.32\linewidth]{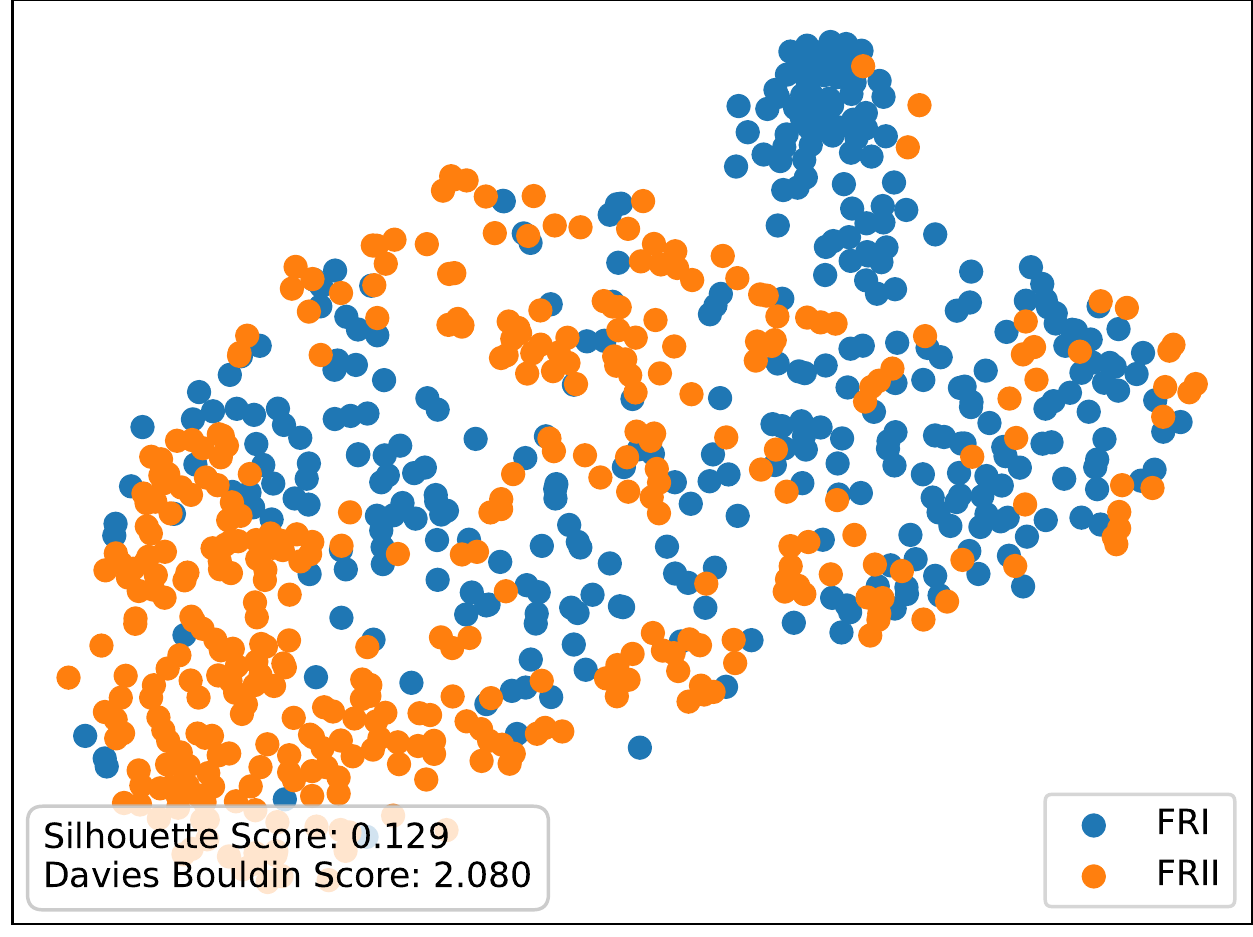}}
    \hfill
    \subfloat[Supervised G-CNN\label{fig:tsne_supervised}]{\includegraphics[width=0.32\linewidth]{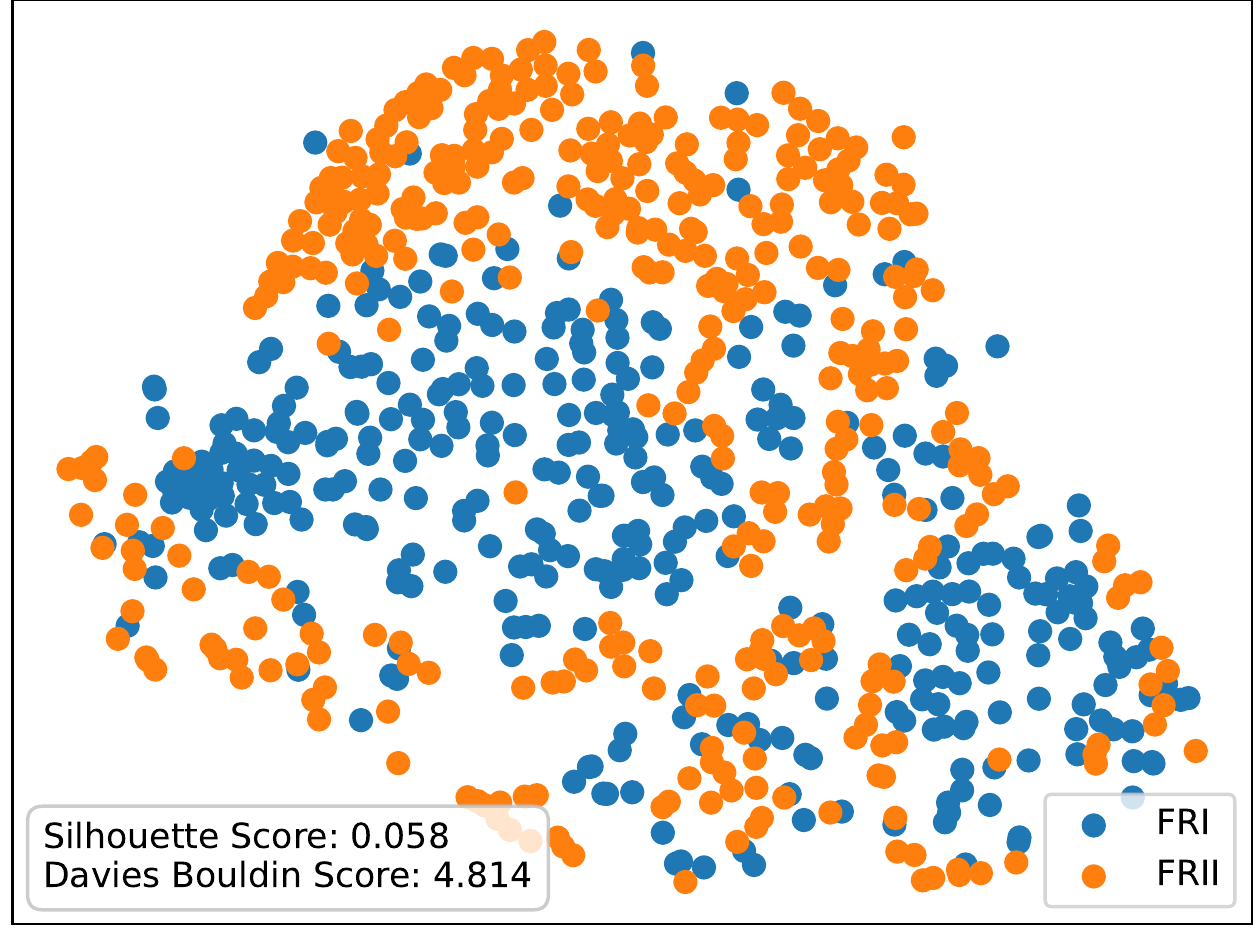}}
    \caption{Visualization of the representations obtained from the first fully connected layer of the fine-tuned encoders of (a) BYOL, (b) SimCLR, and (c) the Supervised G-CNN Model on our labelled dataset using t-SNE. The blue points represent the FRI class, while the orange points represent the FRII class. Our models, as evidenced by the higher Silhouette scores and the lower Davies Bouldin scores, show improved clustering performance, demonstrating that the models learned more effective data representations than the supervised model.}
    \label{fig:cluster_finetuned}
\end{figure*}

\subsection{Experimental setup}\label{setup}
We utilized two self-supervised methods, SimCLR and BYOL, separately on the unlabeled dataset, Dataset-U, to extract the underlying representations of the radio galaxy images.

For the SimCLR model, we fixed the following hyperparameters: (i) learning rate of 0.02, (ii) weight decay of $10^{-6}$, (iii) batch size of 16, (iv) MLP hidden layer dimension of 2048, (v) MLP output layer dimension of 128, and (vi) number of epochs of 500. We used the Layer-wise Adaptive Rate Scaling (LARS) optimizer and Cosine Annealing Warm Restarts scheduler with a minimum learning rate 0.05.

For the BYOL model, we tuned the following hyperparameters: (i) learning rate of $3\times10^{-4}$, (ii) moving average decay of 0.99, (iii) batch size of 16, (iv) MLP hidden layer dimension of 4092, (v) MLP output layer dimension of 256, and (vi) number of epochs of 500. We used the Adam optimizer to train this model.

Once the representations were learned from the unlabeled dataset, we used the learned representations to fine-tune the G-CNN model using a supervised approach on the labelled dataset, Dataset-F. We tuned the following hyperparameters for this step: (i) learning rate of 0.0001, (ii) weight decay of $10^{-6}$, (iii) batch size of 50, and (iv) number of epochs of 600. We used Adam optimizer and a Reduce Learning Rate (LR) on Plateau strategy with patients of 2 and a factor of 0.9.

The experiments in this paper were carried out on a high-performance computing environment equipped with a powerful Intel(R) Core(TM) i9-9900K CPU @ 3.60GHz and 64 Gigabytes of RAM, paired with a cutting-edge GeForce RTX 2080 Ti GPU. The training process for each self-supervised model took approximately 17 hours to complete on this setup, while the downstream training required roughly 40 minutes to run.


\subsection{Result analysis}\label{results}
Several experiments were conducted to compare our method with a traditional supervised approach. Our model's effectiveness is demonstrated through various qualitative and quantitative metrics.

At first, we evaluated the quality of the clusters formed by learned representations with our semi-supervised approaches (BYOL and SimCLR) and compared the results with the supervised model.
To do so, we passed the labeled dataset (Dataset-F) through the fine-tuned encoder for semi-supervised approaches (BYOL,  SimCLR). We extracted the representation from the first fully connected (FC) layer. We have performed the same process for the supervised model. The output of the first FC layer is a 120-dimensional vector, which is impossible to visualize. We employed t-Distributed Stochastic Neighbor Embedding (t-SNE)  to visualize the feature maps in a lower-dimensional space. 
The results of the t-SNE visualization, plotted in Fig. \ref{fig:cluster_finetuned}, indicate that the cluster quality formed by the representations learned from the semi-supervised models is significantly better than those from the supervised model. It means the data points are more clearly clustered.

In order to further validate our observations, we used two widely accepted metrics to quantify the cluster quality: Silhouette Score and Davies Bouldin Score. The Silhouette Score measures the similarity of a sample to its own cluster compared to other clusters, with higher scores indicating better cluster formation. On the other hand, the Davies Bouldin Score measures the average similarity between each cluster and its closest cluster, with lower scores representing better cluster quality.
The results of our evaluation showed that the fine-tuned encoders of semi-supervised BYOL and SimCLR had Silhouette Scores of 0.276 and 0.129, respectively. In contrast, the supervised model had a score of 0.058. The semi-supervised BYOL and SimCLR models also received Davies Bouldin Scores of 1.509 and 2.080, respectively, compared to 4.814 for the supervised model.
These results demonstrate that our semi-supervised models have learned more effective data representations, resulting in significantly improved clustering performance compared to the supervised model.

\begin{table*}
    \caption{Performance comparison between Semi-supervised and Supervised methods; The table illustrates the superiority of our Semi-supervised models over the state-of-the-art supervised method across various classification metrics}
    \label{tab:fr_comp}
    \centering
    \resizebox{0.98\linewidth}{!}{%
    \begin{tabular}{|l|l|lll|lll|}
    \hline
            & & & & & & &   \\[-1.6ex] 
            &    &  \multicolumn{3}{ |c| }{FRI} & \multicolumn{3}{ |c| }{FRII} \\[0.4ex]
            \cline{3-8}
        & & & & & & &   \\[-1.6ex] 
        & Accuracy[\%] & Precision & Recall & f1-score & Precision & Recall & f1-score\\[0.4ex] \hline\hline
        & & & & & & &   \\[-1.6ex] 
       Semi-supervised SimCLR & $\underline{95.77 \pm 0.90}$ & $\boldsymbol{0.98 \pm 0.061}$ & $0.93 \pm 0.018$ & $\underline{0.95 \pm 0.011}$ & $0.94 \pm 0.013$ & $\boldsymbol{0.98 \pm 0.014}$ & $\underline{0.96 \pm 0.009}$ \\
       & & & & & & &   \\[-1.6ex] 
       Semi-supervised BYOL & $\boldsymbol{97.12 \pm 0.40}$ & $\underline{0.97 \pm 0.008}$ & $\underline{0.96 \pm 0.009}$ & $\boldsymbol{0.97 \pm 0.005}$ & $\underline{0.96 \pm 0.007}$ & $\underline{0.98 \pm 0.008}$ & $\boldsymbol{0.97 \pm 0.004}$ \\[0.4ex] \hline
         & & & & & & &   \\[-1.6ex] 
        Supervised G-CNN & $94.80 \pm 0.90$ & $0.93 \pm 0.012$ & $\boldsymbol{0.96 \pm 0.010}$ & $0.94 \pm 0.009$ & $\boldsymbol{0.96 \pm 0.009}$ & $0.94 \pm 0.012$ & $0.95 \pm 0.009$ \\[0.4ex] \hline
    \end{tabular}}
\end{table*}

\begin{figure}
    \centering
    \subfloat[Training Loss\label{fig:convergence_train}]{\includegraphics[width=0.99\linewidth]{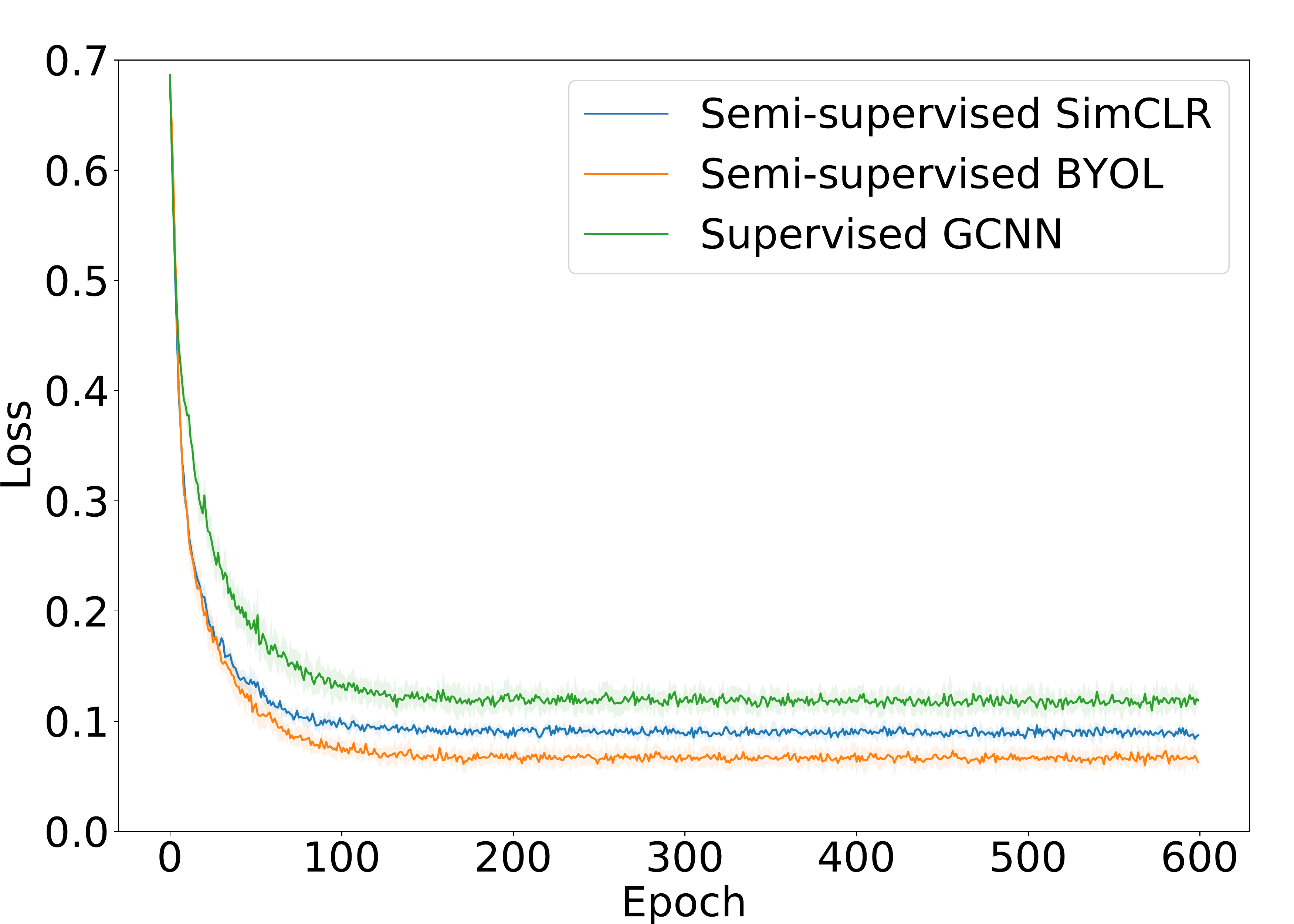}}
    \hfill
    \subfloat[Validation Loss\label{fig:convergence_val}]{\includegraphics[width=0.99\linewidth]{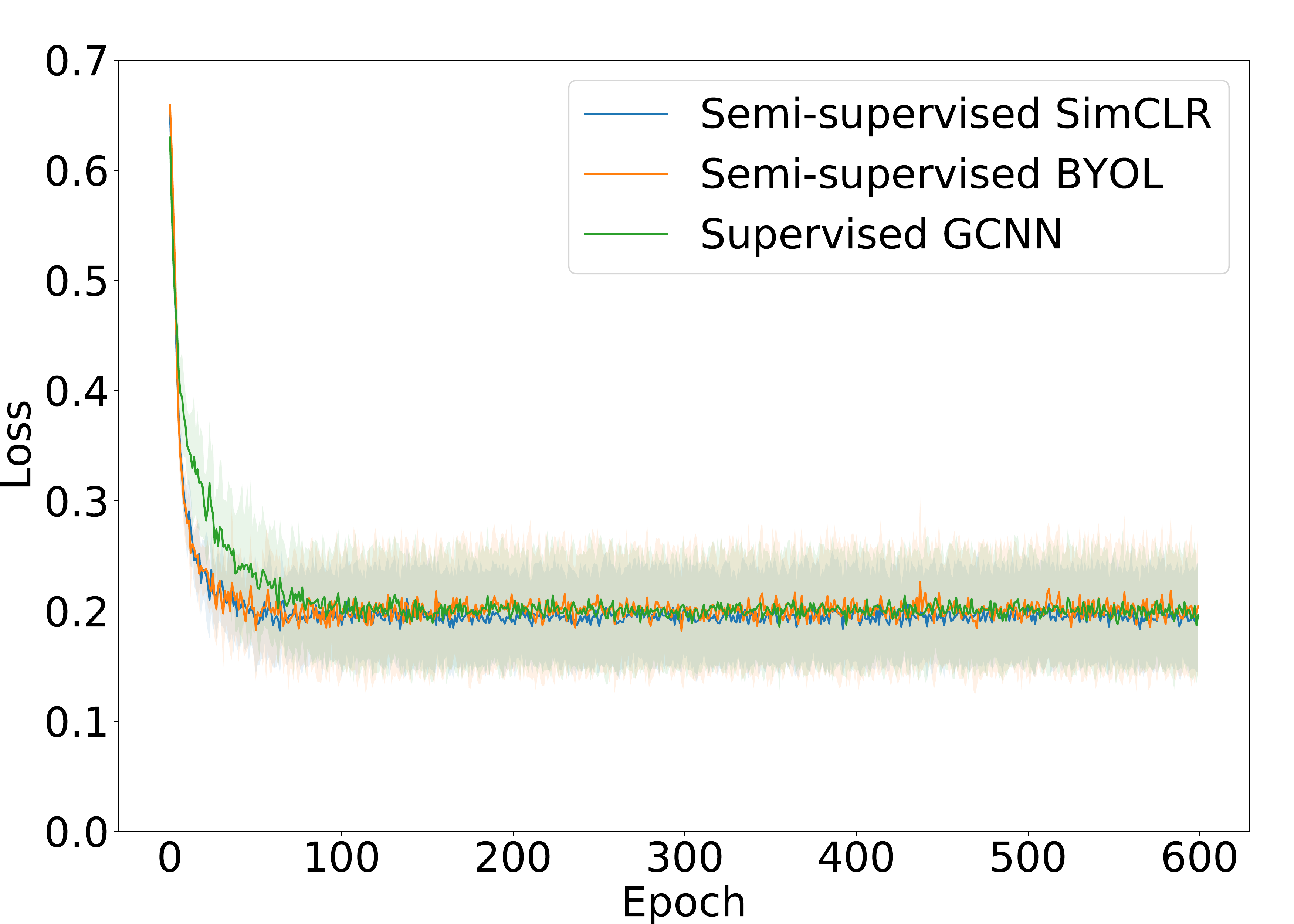}}

    \caption{Convergence plots for fine-tuning the encoders of SimCLR and BYOL, compared to training a supervised G-CNN on Dataset-F. The plots show the mean and standard deviation of (a) training loss and (b) validation loss across 5-fold cross-validation. The fast convergence of the fine-tuned encoders compared to the supervised G-CNN suggests the effectiveness of the learned representations.}
    \label{fig:convergence}
\end{figure}

In order to evaluate the convergence of our models during the fine-tuning stage, we performed 5-fold cross-validation on the labelled dataset (Dataset-F) using the representations learned by the encoders of SimCLR and BYOL. 
We also compared the convergence of the loss function in the fine-tuning step with that of the supervised G-CNN approach. For each repetition, we used a different fold for validation and the remaining four for training. 
Fig. \ref{fig:convergence_train} and \ref{fig:convergence_val} show the convergence plots for the training and validation loss, respectively. 
The plots were generated by taking the mean and standard deviation of the loss across the five repeats. As seen from the plots, our fine-tuning approach's training and validation losses converge very quickly compared to the supervised G-CNN approach, indicating that the representations learned in the first stage are adequate for the downstream task.

In order to further evaluate the performance of our semi-supervised models, we fine-tuned the encoders of SimCLR and BYOL using a labelled dataset (Dataset-F). We compared the results to a state-of-the-art supervised model, D16-Steerable G-CNN.
We trained and tested all models using the same setup and repeated the process 15 times for each model.
As shown in Table \ref{tab:fr_comp}, the results include the average precision, recall, and F1 scores for both FRI and FRII classes, along with the standard deviation.
The best metrics are highlighted in bold, and the second-best is underlined.

The results demonstrate that our semi-supervised models outperform the supervised model.
Both BYOL and SimCLR achieved higher accuracy than D16-Streeable G-CNN. 
Semi-supervised BYOL achieved the highest accuracy of $97.12\%$, followed by SimCLR at $95.77\%$. 
In addition, both models exhibited superior performance in other classification metrics.
For example, SimCLR achieved precision and recall of $98\%$ for the FRI and FR II radio galaxies, respectively, and BYOL achieved an excellent f-score of $97\%$ for both classes.
These results demonstrate the effectiveness of our proposed semi-supervised models in accurately classifying radio galaxies.

In addition to the results obtained from the classification metrics, we further evaluated the performance of our models using Receiver Operating Characteristic (ROC) curves and the corresponding Area Under the Curve (AUC) scores. 
The ROC curve is a graphical representation of the performance of a binary classifier system, while the AUC score provides a single numeric value to represent the classifier's performance.
It ranges from 0 to 1, with a score of 1 indicating a perfect classifier and a score of 0.5 indicating a classifier that performs no better than random.

\begin{figure}
    \centering
    \includegraphics[width=0.98\linewidth]{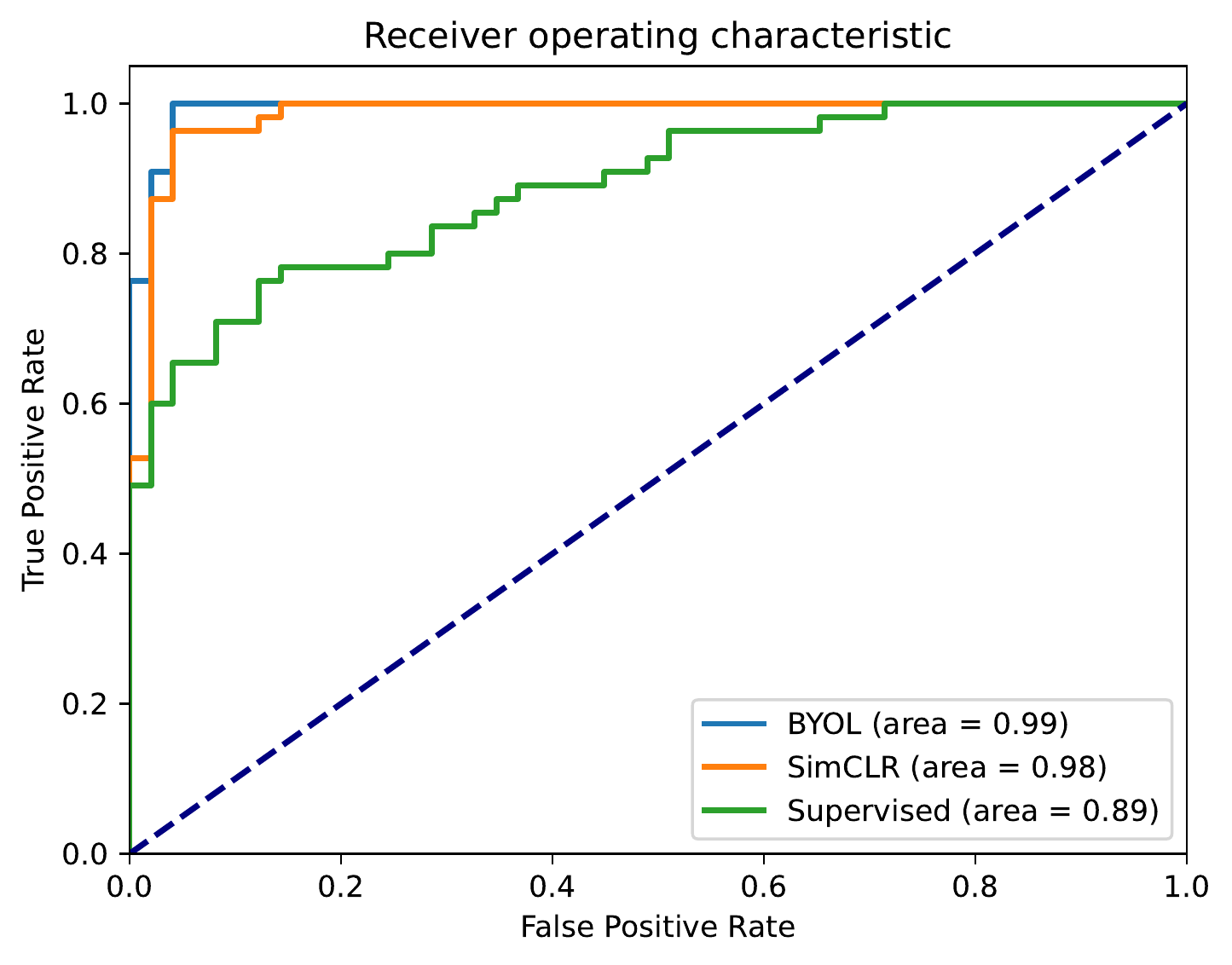}
    \caption{Receiver Operating Characteristic (ROC) curves for the fine-tuned encoders of BYOL, SimCLR and the Supervised model, displaying the performance of the models in classifying the two classes of radio galaxies. The corresponding Area Under the Curve (AUC) scores are also shown, with a score of 1 indicating a perfect classifier and a score of 0.5 indicating a classifier that performs no better than random.}
    \label{fig:roc}
\end{figure}

The evaluation results are depicted in Fig. \ref{fig:roc}. The AUC scores for semi-supervised BYOL and SimCLR's fine-tuned encoders were found to be 0.99 and 0.98, respectively, while the score for the Supervised model was 0.89. These high AUC scores for BYOL and SimCLR indicate their exceptional ability to distinguish between the two classes of radio galaxies accurately. The models can effectively differentiate between radio galaxies with a minimal rate of false positives, which implies that they can correctly identify most radio galaxies without misclassifying them. 

Overall, these results support the effectiveness of our proposed semi-supervised method in classifying radio galaxies. The high AUC scores and the shape of the ROC curves indicate that our models can effectively differentiate between the two classes of radio galaxies. This makes them valuable for further analysis and studies in the field.

To further demonstrate the superiority of our proposed models over the supervised approach, we conducted a statistical t-test on the accuracy scores of 15 runs of each model.
The paired t-test between the supervised G-CNN and semi-supervised SimCLR models yielded a t-value of approximately $-1.89$ and a p-value of approximately $0.08$. Similarly, the test between the supervised G-CNN and semi-supervised BYOL models yielded a t-value of approximately $-3.47$ and a p-value of approximately $0.0038$. These results indicate that our proposed semi-supervised method performs significantly better than the supervised G-CNN, with a high level of statistical significance.


\section{Conclusion}\label{conclusion}

This paper presents a novel approach for classifying radio galaxies with considerable effectiveness.
Using a vast amount of unlabeled images during the pre-training stage allowed our models to learn robust and invariant features.
The group-equivariant architecture is vital in enhancing the model's capacity to extract structural and symmetry information, which is crucial for accurate classification.

Fine-tuning the pre-trained encoder with a limited amount of labelled data significantly improved its performance compared to traditional supervised methods.
Our quantitative analysis, using Silhouette Score and Davies Bouldin Score, showed that the fine-tuned encoders produced higher Silhouette scores and lower Davies Bouldin scores, indicating better cluster formation compared to the supervised model.
The visualization of the representations, obtained from the first fully connected layer of the fine-tuned encoders, demonstrated a clear separation between FRI and FRII galaxies into different clusters. 
This highlights the effectiveness of our task-agnostic self-supervised learning approach in utilizing limited labelled data to learn effective data representations.

Our semi-supervised method significantly improved over traditional supervised methods, achieving a high accuracy of $97.12\%$ using the fine-tuned BYOL encoder compared to only $94.80\%$ for the supervised approach.
Furthermore, our training and validation losses converged much faster than in the supervised approach, and the ROC curves and AUC score confirmed the ability of our models to distinguish between the two classes of radio galaxies effectively.

To further solidify our findings, a paired t-test on the accuracy score of 15 runs showed a statistically significant improvement in our proposed models over the supervised model.
These results highlight the immense potential of semi-supervised learning in classifying radio galaxies and demonstrate the importance of utilizing vast amounts of high-quality images.

Our work aims to contribute to the classification of the many radio galaxies yet to be discovered, and we hope it encourages further research to improve classification accuracy. While our method shows promise in this context, further studies are required to determine its applicability to other classification tasks with limited labelled data.

With the arrival of upcoming radio telescopes such as SKA, we anticipate a significant increase in the number of high-quality images available for astronomical object classification. Despite its limitations, our method represents an attempt to provide a scalable and efficient solution to this challenge.


\section{Acknowledgment}\label{acknowledgment}
This project has been jointly sponsored by Independent University, Bangladesh and the ICT Division of the Bangladesh Government.


\bibliographystyle{IEEEtran}
\bibliography{IEEEabrv,IJCNN}

\begin{thebibliography}{10}
\providecommand{\url}[1]{#1}
\csname url@samestyle\endcsname
\providecommand{\newblock}{\relax}
\providecommand{\bibinfo}[2]{#2}
\providecommand{\BIBentrySTDinterwordspacing}{\spaceskip=0pt\relax}
\providecommand{\BIBentryALTinterwordstretchfactor}{4}
\providecommand{\BIBentryALTinterwordspacing}{\spaceskip=\fontdimen2\font plus
\BIBentryALTinterwordstretchfactor\fontdimen3\font minus
  \fontdimen4\font\relax}
\providecommand{\BIBforeignlanguage}[2]{{%
\expandafter\ifx\csname l@#1\endcsname\relax
\typeout{** WARNING: IEEEtran.bst: No hyphenation pattern has been}%
\typeout{** loaded for the language `#1'. Using the pattern for}%
\typeout{** the default language instead.}%
\else
\language=\csname l@#1\endcsname
\fi
#2}}
\providecommand{\BIBdecl}{\relax}
\BIBdecl

\bibitem{fanaroff74}
B.~L. {Fanaroff} and J.~M. {Riley}, ``{The morphology of extragalactic radio
  sources of high and low luminosity},'' \emph{\mnras}, vol. 167, pp. 31P--36P,
  May 1974.

\bibitem{sasmal2022}
T.~K. {Sasmal}, S.~{Bera}, S.~{Pal}, and S.~{Mondal}, ``{A New Catalog of
  Head-Tail Radio Galaxies from the VLA FIRST Survey},'' \emph{\apjs}, vol.
  259, no.~2, p.~31, Apr. 2022.

\bibitem{proctor11}
D.~D. {Proctor}, ``{Morphological Annotations for Groups in the First
  Database},'' \emph{\apjs}, vol. 194, no.~2, p.~31, Jun. 2011.

\bibitem{leahy92}
J.~P. {Leahy} and P.~{Parma}, ``{Multiple outbursts in radio galaxies.}'' in
  \emph{Extragalactic Radio Sources. From Beams to Jets}, J.~{Roland},
  H.~{Sol}, and G.~{Pelletier}, Eds., Jan. 1992, pp. 307--308.

\bibitem{chen20}
T.~Chen, S.~Kornblith, M.~Norouzi, and G.~Hinton, ``A simple framework for
  contrastive learning of visual representations,'' in \emph{International
  conference on machine learning}.\hskip 1em plus 0.5em minus 0.4em\relax PMLR,
  2020, pp. 1597--1607.

\bibitem{grill20}
J.-B. {Grill}, F.~{Strub}, F.~{Altch{\'e}}, C.~{Tallec}, P.~H. {Richemond},
  E.~{Buchatskaya}, C.~{Doersch}, B.~{Avila Pires}, Z.~D. {Guo}, M.~{Gheshlaghi
  Azar}, B.~{Piot}, K.~{Kavukcuoglu}, R.~{Munos}, and M.~{Valko}, ``{Bootstrap
  your own latent: A new approach to self-supervised Learning},'' \emph{arXiv
  e-prints}, p. arXiv:2006.07733, Jun. 2020.

\bibitem{scaife21}
A.~M.~M. {Scaife} and F.~{Porter}, ``{Fanaroff-Riley classification of radio
  galaxies using group-equivariant convolutional neural networks},''
  \emph{\mnras}, vol. 503, no.~2, pp. 2369--2379, May 2021.

\bibitem{miraghaei17}
H.~{Miraghaei} and P.~N. {Best}, ``{The nuclear properties and extended
  morphologies of powerful radio galaxies: the roles of host galaxy and
  environment},'' \emph{\mnras}, vol. 466, no.~4, pp. 4346--4363, Apr. 2017.

\bibitem{galvin20}
T.~J. {Galvin}, M.~T. {Huynh}, R.~P. {Norris}, X.~R. {Wang}, E.~{Hopkins},
  K.~{Polsterer}, N.~O. {Ralph}, A.~N. {O'Brien}, and G.~H. {Heald},
  ``{Cataloguing the radio-sky with unsupervised machine learning: a new
  approach for the SKA era},'' \emph{\mnras}, vol. 497, no.~3, pp. 2730--2758,
  Sep. 2020.

\bibitem{aniyan17}
A.~K. {Aniyan} and K.~{Thorat}, ``{Classifying Radio Galaxies with the
  Convolutional Neural Network},'' \emph{\apjs}, vol. 230, no.~2, p.~20, Jun.
  2017.

\bibitem{alhassan18}
W.~{Alhassan}, A.~R. {Taylor}, and M.~{Vaccari}, ``{The FIRST Classifier:
  compact and extended radio galaxy classification using deep Convolutional
  Neural Networks},'' \emph{\mnras}, vol. 480, no.~2, pp. 2085--2093, Oct.
  2018.

\bibitem{wu19}
C.~{Wu}, O.~I. {Wong}, L.~{Rudnick}, S.~S. {Shabala}, M.~J. {Alger}, J.~K.
  {Banfield}, C.~S. {Ong}, S.~V. {White}, A.~F. {Garon}, R.~P. {Norris},
  H.~{Andernach}, J.~{Tate}, V.~{Lukic}, H.~{Tang}, K.~{Schawinski}, and F.~I.
  {Diakogiannis}, ``{Radio Galaxy Zoo: CLARAN - a deep learning classifier for
  radio morphologies},'' \emph{\mnras}, vol. 482, no.~1, pp. 1211--1230, Jan.
  2019.

\bibitem{tang19}
H.~{Tang}, A.~M.~M. {Scaife}, and J.~P. {Leahy}, ``{Transfer learning for radio
  galaxy classification},'' \emph{\mnras}, vol. 488, no.~3, pp. 3358--3375,
  Sep. 2019.

\bibitem{cohen16}
T.~Cohen and M.~Welling, ``Group equivariant convolutional networks,'' in
  \emph{International conference on machine learning}.\hskip 1em plus 0.5em
  minus 0.4em\relax PMLR, 2016, pp. 2990--2999.

\bibitem{ma19}
Z.~{Ma}, H.~{Xu}, J.~{Zhu}, D.~{Hu}, W.~{Li}, C.~{Shan}, Z.~{Zhu}, L.~{Gu},
  J.~{Li}, C.~{Liu}, and X.~{Wu}, ``{A Machine Learning Based Morphological
  Classification of 14,245 Radio AGNs Selected from the Best-Heckman Sample},''
  \emph{\apjs}, vol. 240, no.~2, p.~34, Feb. 2019.

\bibitem{ma19c}
Z.~{Ma}, J.~{Zhu}, and H.~{Zhu}, Yongkai~{Xu}, ``{Classification of Radio
  Galaxy Images with Semi-supervised Learning},'' in \emph{Data Mining and Big
  Data}, ser. Communications in Computer and Information Science, {{Tan}, Ying
  and {Shi}, Yuhui}, Ed., vol. 1071.\hskip 1em plus 0.5em minus 0.4em\relax 4th
  International Conference, DMBD 2019, Chiang Mai, Thailand: Springer, July
  2019, pp. 191--200.

\bibitem{slijepcevic22}
I.~V. {Slijepcevic}, A.~M.~M. {Scaife}, M.~{Walmsley}, M.~{Bowles}, O.~I.
  {Wong}, S.~S. {Shabala}, and H.~{Tang}, ``{Radio Galaxy Zoo: using
  semi-supervised learning to leverage large unlabelled data sets for radio
  galaxy classification under data set shift},'' \emph{\mnras}, vol. 514,
  no.~2, pp. 2599--2613, Aug. 2022.

\bibitem{banfield15}
J.~K. {Banfield}, O.~I. {Wong}, K.~W. {Willett}, R.~P. {Norris}, L.~{Rudnick},
  S.~S. {Shabala}, B.~D. {Simmons}, C.~{Snyder}, A.~{Garon}, N.~{Seymour},
  E.~{Middelberg}, H.~{Andernach}, C.~J. {Lintott}, K.~{Jacob}, A.~D.
  {Kapi{\'n}ska}, M.~Y. {Mao}, K.~L. {Masters}, M.~J. {Jarvis},
  K.~{Schawinski}, E.~{Paget}, R.~{Simpson}, H.~R. {Kl{\"o}ckner},
  S.~{Bamford}, T.~{Burchell}, K.~E. {Chow}, G.~{Cotter}, L.~{Fortson},
  I.~{Heywood}, T.~W. {Jones}, S.~{Kaviraj}, {\'A}.~R. {L{\'o}pez-S{\'a}nchez},
  W.~P. {Maksym}, K.~{Polsterer}, K.~{Borden}, R.~P. {Hollow}, and L.~{Whyte},
  ``{Radio Galaxy Zoo: host galaxies and radio morphologies derived from visual
  inspection},'' \emph{\mnras}, vol. 453, no.~3, pp. 2326--2340, Nov. 2015.

\bibitem{porter20}
\BIBentryALTinterwordspacing
F.~A.~M. Porter, ``Mirabest batched dataset,'' Nov. 2020. [Online]. Available:
  \url{https://doi.org/10.5281/zenodo.4288837}
\BIBentrySTDinterwordspacing

\end{thebibliography}

\end{document}